\documentclass[%
 reprint,
 amsmath,amssymb,
 aps,
]{revtex4-1}

\usepackage{graphicx}
\usepackage{dcolumn}
\usepackage{bm}

\begin{document}


\title{Nonreciprocal photon blockade in a two-mode cavity with a second-order nonlinearity}

\author{Kai Wang}
\author{Ya-Fei Yu}
\author{Zhi-Ming Zhang}%
 \email{zmzhang@scnu.edu.cn}
\affiliation{%
 Guangdong Provincial Key Laboratory of Nanophotonic Functional Materials and Devices (School of
Information and Optoelectronic Science and Engineering), and Guangdong Provincial Key Laboratory of
Quantum Engineering and Quantum Materials, South China Normal University, Guangzhou 510006, China}%

\date{\today}

\begin{abstract}
It is shown that the Fizeau drag can be used to cause nonreciprocity. We propose the use of a nanostructured toroid cavity made of $\chi^{(2)}$ nonlinear materials to achieve nonreciprocal photon blockade (PB) through the Fizeau drag. Under the weak driving condition, we discuss the origins of the PB based on the doubly resonant modes with good spatial overlap at the fundamental and second-harmonic frequencies. We also find that for the fundamental mode, the PB happens when we drive the system from one side but the photon-induced tunneling happens when we drive the system from the other side. However, there is no such phenomenon in the second-harmonic mode. Remarkably, the PB phenomenon occurs with a reasonably small optical nonlinearity thus bringing the system parameters closer to the reasonably achievable realm by the current technology.
\end{abstract}

\pacs{Valid PACS appear here}
\maketitle

\section{\label{sec:level1}Introduction}
Considerable efforts have been dedicated to the study of  achieving photon blockade (PB), which is a quantum phenomenon that can be exploited to convert a coherent classical light source of defined wavelength into antibunched photon streams \cite{PhysRevA.61.011801}. It is indispensable in a variety of practical applications. Quantum information processing \cite{PhysRevLett.109.013603,10.1038/35005001,Buluta_2011} and quantum cryptography \cite{RevModPhys.81.1301,10.1038/nphoton.2009.229,10.1038/nature07127} are just a few of the most widely known examples. There have been various studies on how to produce single photons in circuit-QED systems \cite{PhysRevLett.106.243601,PhysRevLett.107.053602,PhysRevA.89.043818}, cavity QED systems \cite{PhysRevLett.94.053604,PhysRevLett.96.093603}, optomechanical systems \cite{PhysRevLett.107.063601,PhysRevA.92.033806}, coupled cavities \cite{PhysRevA.76.031805,PhysRevA.94.013815,PhysRevA.91.063808} and cavity-free systems \cite{PhysRevLett.107.223601,10.1038/nature11361}. In these pioneering studies, the PB is generated in weakly-nonlinear systems which allow for destructive quantum interference between distinct driven-dissipative
pathways \cite{PhysRevA.96.053810}, called unconventional PB or arises from the anharmonicity in energy eigenvalues of the systems caused by strong nonlinearity \cite{PhysRevLett.121.153601}. Although the unconventional PB requires a significant smaller optical nonlinearity than its conventional counterpart, it can exhibit higher-order bunched photons. Thus the unconventional PB is not a good way of generating single photons in general.

Nonreciprocal devices, which break the physical symmetry, allowing light propagating from one side but not the other, are also playing a very important role in a wide range of applications, such as signal processing and invisible sensing \cite{Sounas2017}. It is typically achieved in previous experiments on the classical regimes based on atomic gases \cite{PhysRevLett.120.043901,PhysRevLett.110.093901}, nonlinear optics \cite{Fan447,PhysRevLett.118.033901}, optomechanics \cite{PhysRevLett.102.213903,10.1038/nphoton.2016.161} and non-Hermitian optics  \cite{PhysRevLett.110.234101,nphoton.2014.133} as well as studied on the quantum regimes based on rotating resonators \cite{PhysRevLett.121.153601}.  

In this paper, We explore the possibility of achieving nonreciprocal PB in a rotating toroid
cavity via $\chi^{(2)}$ susceptibility, following the proposal in Ref.\cite{PhysRevLett.121.153601} which relies on the conventional condition of strong nonlinearity to obtain nonreciprocal PB in a spinning Kerr resonator. Under the weak driving condition, we compare the difference between the origins of PB phenomena in both modes and investigate the nonreciprocity occurs in the fundamental mode. Note that the present PB devices, relying on the $\chi ^{(2)}$ nonlinearity, can potentially be achieved with larger values of the nonlinear interaction compared with the devices relying on the Kerr-type
nonlinearity \cite{PhysRevA.89.031803}. This will eventually bring the system parameters closer to the reasonably achievable realm by the current technology.

The remainder of this paper is organized as follows: In Sec.\ref{sec:level2}, we propose a physical model and analytically study the PB and the nonreciprocity phenomena. In Sec.\ref{sec:level3}, we numerically examine the analysis in Sec.\ref{sec:level2} by investigating the photon statistical properties via the quantum master equation. Finally, we summarize the work in Sec.\ref{sec:level4}.

\section{\label{sec:level2}Model and theory}
\begin{figure}[! htbp]
    \centering\includegraphics[width=0.45\textwidth]{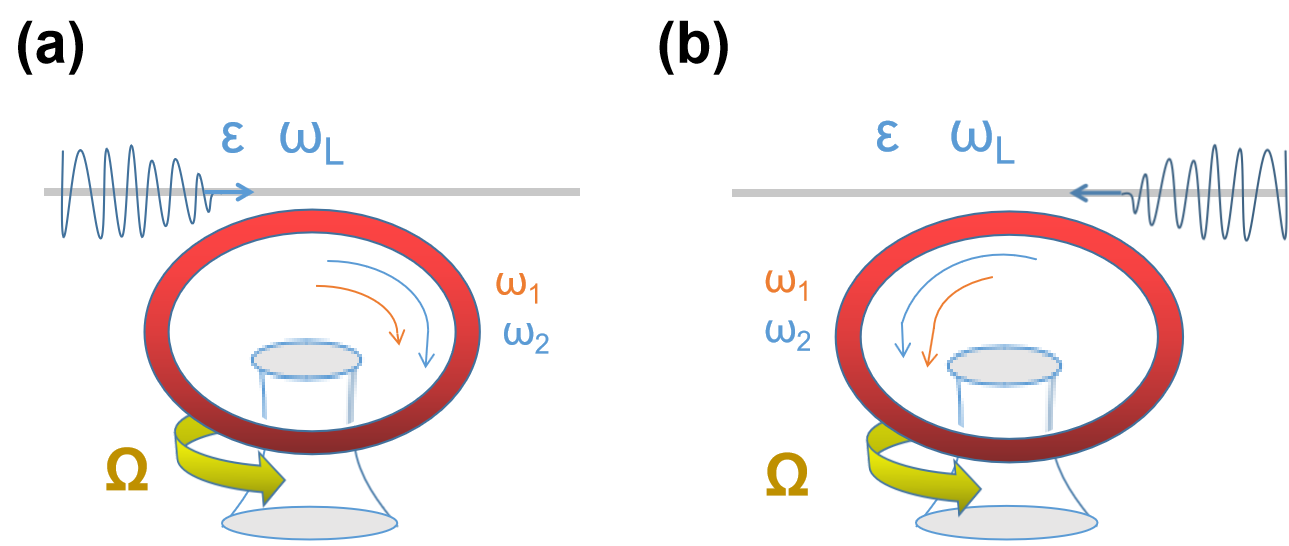}
    \caption{A rotating optical cavity with a $\chi^{ (2)}$ nonlinear material. The cavity is rotating at a fixed angular velocity $\Omega$ and an external classical light is coupled into and out of the cavity through an optical fiber. (a) $\Delta _F >0$ when we drive the device from its left side. (b) $\Delta _F <0$ when we drive the device from its right side. }
    \label{fig.1}
\end{figure}

We consider a rotating optical cavity based on doubly resonant modes with good spatial overlap at the fundamental and second-harmonic frequencies via a $\chi^{ (2)}$ nonlinear material as shown in Fig.\ref{fig.1}(a)(b).
The cavity is rotating with a fixed angular velocity $\Omega$ and an external classical light is coupled into and out of the cavity through a optical fiber. Therefore for a $\chi^{ (2)}$ nonlinear material, the rotating cavity Hamiltonian is given by\cite{PhysRevLett.96.057405}
\begin{equation}
\begin{aligned}
H&=\hbar (\omega _1 +\Delta_{F_1}) a^ \dagger a+\hbar (\omega _2+\Delta_{F_2}) b^ \dagger b\\
& \quad +\hbar g (b {a^ \dagger}^2+ b ^ \dagger a^2)\\
\label{1}
\end{aligned}
\end{equation}
where $\omega _1$ and $\omega _2$ are the frequencies of the quantized fundamental mode and the quantized second-harmonic mode, respectively. Both of the cavity modes experience a Fizeau shift because of the rotating, hence we have $\omega _1 \rightarrow \omega_1 + \Delta _{F_1}$ and $\omega_2 \rightarrow \omega _2 +\Delta _{F_2}$. $\Delta _{F_1}$ and $\Delta _{F_2} $ here are decided by the fixed angular velocity of the rotating cavity according to \cite{Malykin_2000}
\begin{equation}
    \Delta_{F_i}=\pm \frac{nr \Omega \omega _i}{c}(1-\frac{1}{n^2}-\frac{\lambda}{n}\frac{dn}{d \lambda}),(i=1,2)
    \label{2}
\end{equation}
where $n$ is the refractive index, $r$ is the cavity radius,
$c$ is the speed of light in vacuum and $\lambda$ is the wavelength of the external classical light. Here, $\Delta _{F_1}, \Delta _{F_2} >0$ $(\Delta _{F_1}, \Delta _{F_2}<0)$ denotes that the light propagating against (along) the direction of the rotating cavity as shown in Fig.\ref{fig.1}a  (Fig.\ref{fig.1}b), i.e., driving the device from its left (right) side in our case. $\omega_2 =2 \omega _1$ is considered in this paper for simplicity and we use $\Delta_{F}$ to express $\Delta_{F_1}$ for below convenience, therefore we have $\Delta_{F_2}=2\Delta_F$ since $\omega_2 =2 \omega _1$ was assumed. $a(a^ \dagger)$ and $b(b^\dagger)$ represent annihilation (creation) operators for the fundamental mode and the second-harmonic mode respectively and $g$ is the hopping interaction between the two modes which is proportional to $\chi ^{(2)}$\cite{PhysRevB.87.235319}. Here we notice that the last term with $g$ leads to an anharmonic energy-level structure.

In the following, we study the PB phenomenon in the system arising as a result of this two-mode cavity Hamiltonian.
\begin{figure}[! htbp]
    \centering\includegraphics[width=0.45\textwidth]{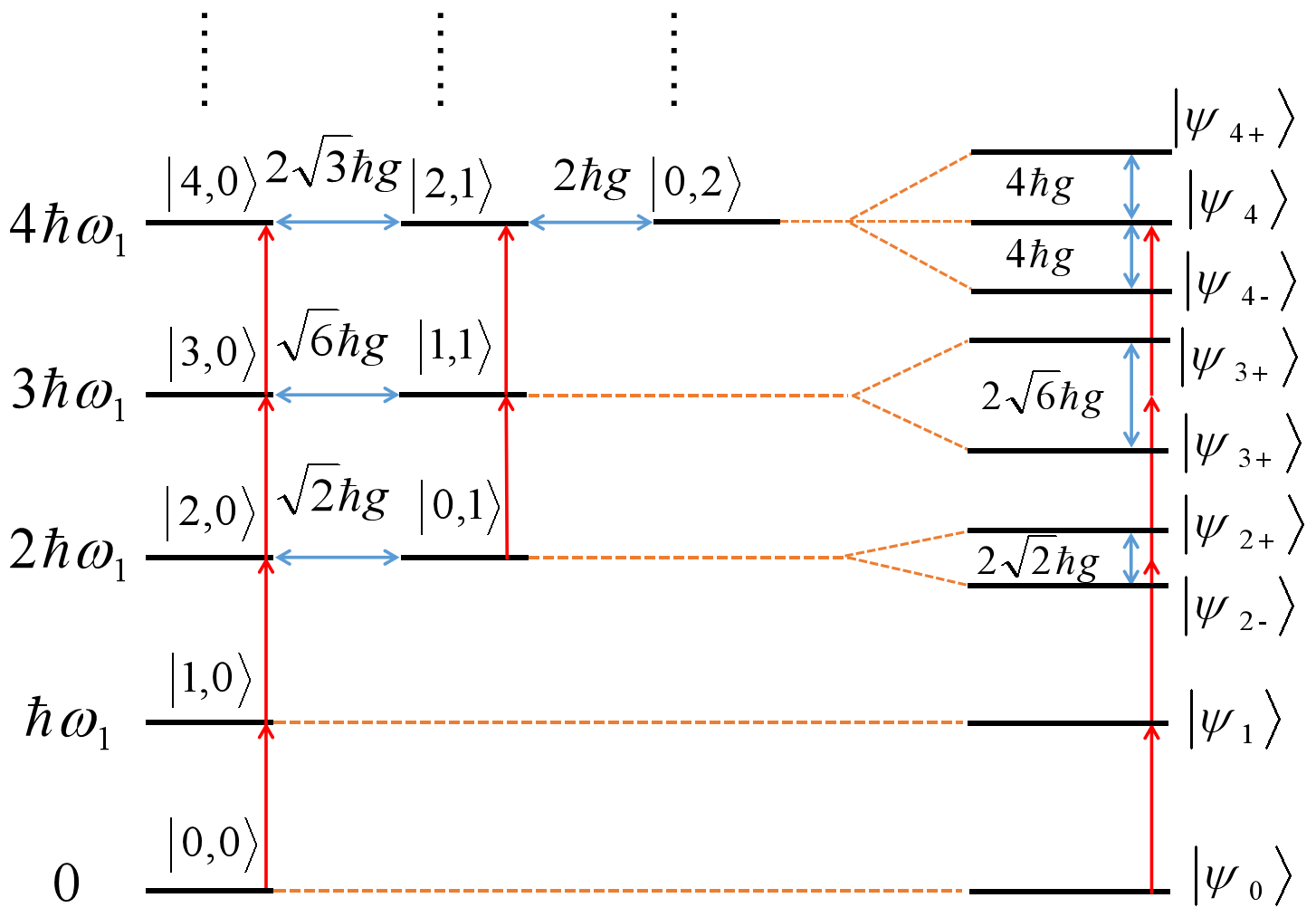}
    \caption{The schematic energy spectrum of the nonlinear cavity based on the $\chi^{(2)} $ material, with the noncoupling Fock state (left) and energy eigenstates (right) when the angular velocity $\Omega=0$. The red arrows show the frequency of the driving laser.}
    \label{fig.2}
\end{figure}
\begin{figure}[! htbp]
    \centering\includegraphics[width=0.4\textwidth]{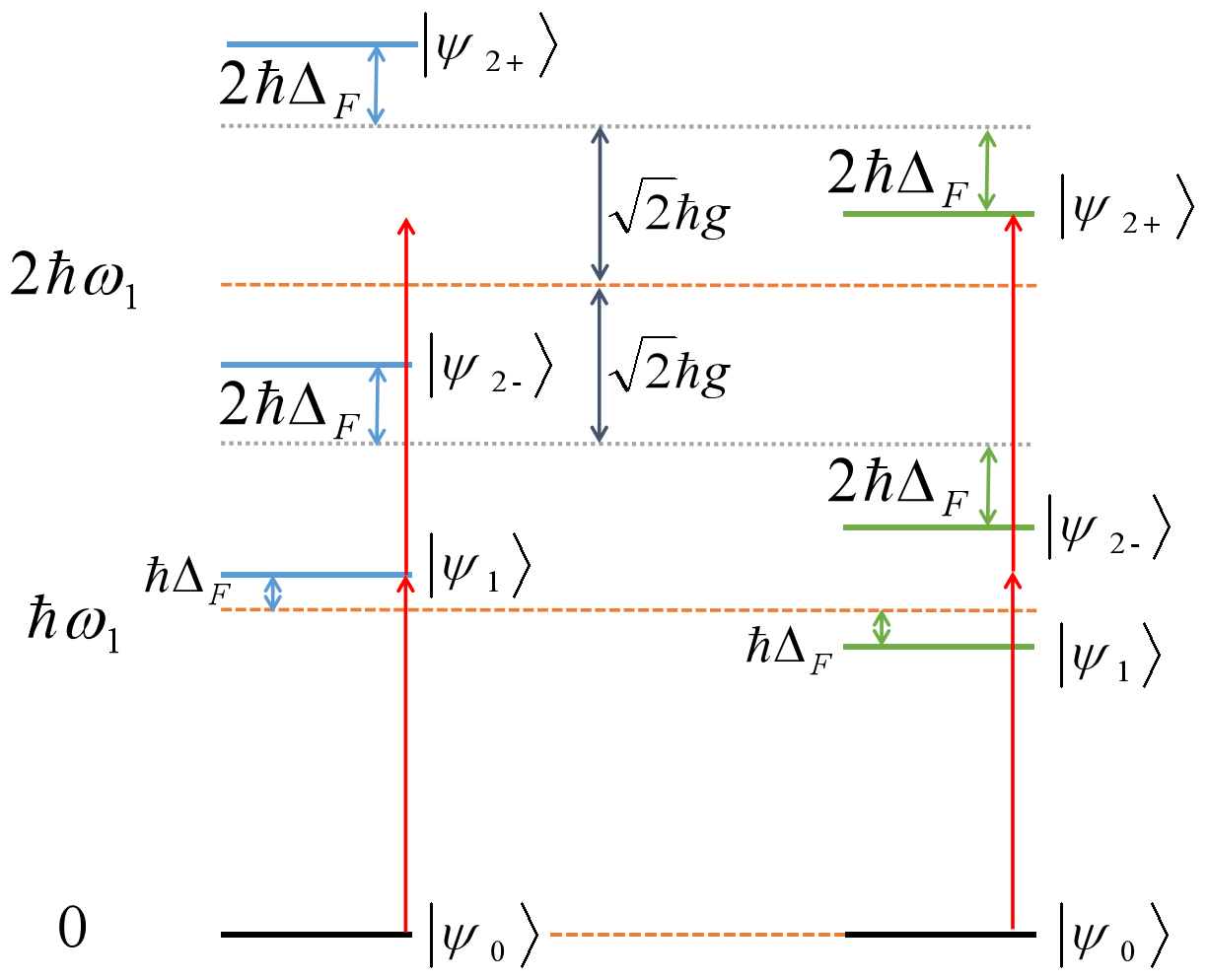}
    \caption{The schematic energy-level transitions of the eigenstates for the doubly resonant nonlinear cavity caused by driving the system from the left side (left) and the right side (right). The angular velocity $\Omega$ satisfies $\Delta_{F}=\pm \frac{nr \Omega \omega _1}{c}(1-\frac{1}{n^2}-\frac{\lambda}{n}\frac{dn}{d \lambda})=\pm \frac{\sqrt{2}}{4}g$. The red arrows show the frequency of the driving laser. }
    \label{fig.3}
\end{figure}
Firstly, we consider the eigen-equation of this system which can be written as $H|\psi _k \rangle =E_k|\psi _k \rangle $ when the fixed angular velocity $\Omega=0$. The eigenstates are $|\psi _k \rangle $ which can be written in terms of the Fock state $|n_a ,n_b\rangle $, where $n_a$ and $n_b$ are the numbers of the photons in the fundamental and second-harmonic modes, and the eigenenergies are $E_k$ ($k=1,2,3 \cdots $).  The energy spectrum is sketched in the diagram as shown in Fig.\ref{fig.2}, with the noncoupling Fock state (left) and energy eigenstates (right). We can see that in the zero-excitation subspace, we have $H|\psi _0 \rangle =E_0|\psi _0 \rangle $, with the eigenstate $|\psi _0 \rangle =|0,0\rangle $ and the eigenvalue $E_0=0$. In the single-excitation subspace, we have $H|\psi _1 \rangle =E_1|\psi _1 \rangle$, with the eigenstate $|\psi _1 \rangle =|1,0\rangle $ and the eigenvalue $E_1=\hbar \omega_1$. However, in the second-excitation subspace, we have $H|\psi _{2 \pm} \rangle =E_{2 \pm}|\psi _{2 \pm} \rangle$, where states $|0,1\rangle $ and $|2,0\rangle $ mix together giving the eigenstates $|\psi _{2 -} \rangle =\frac{\sqrt{2}}{2}(|0 , 1\rangle -|2, 0\rangle ) $ and $|\psi _{2 +} \rangle =\frac{\sqrt{2}}{2}(|0, 1\rangle +|2, 0\rangle )$ with the eigenvalues $E_{2 -}=2\hbar \omega_1- \sqrt{2} \hbar g$ and $E_{2 +}=2\hbar \omega_1+ \sqrt{2} \hbar g$. Therefore, when the light is resonantly coupled to the transition $|\psi _0 \rangle \rightarrow |\psi _1 \rangle  $, the transition $|\psi _1 \rangle \rightarrow |\psi _{2 \pm}\rangle$ is detuned by $\pm \sqrt{2} \hbar g$ and, thus, be suppressed. It means for the fundamental mode, once a photon is coupled into the cavity, it suppresses the occurrence of the second photon with the same frequency going into the cavity, as there is no available state. This condition holds as long as the driving power remains weak, since higher than the  third-excitation subspace, there will be other eigenstates resonant with the driving laser such as $|\psi _4 \rangle$ as we can see in Fig.\ref{fig.2}, which will effectively destroy the PB effect. We can also see that there are two paths for generating two photons in the second-harmonic mode, i.e., $|4,0 \rangle \rightarrow |2,1 \rangle \rightarrow |0,2 \rangle$ and $|1,1 \rangle \rightarrow |2,1 \rangle \rightarrow |0,2 \rangle$, which according to \cite{PhysRevA.96.053810}, will arise unconventional photon blockade in the second-harmonic mode. Then we consider the energy-level transitions caused by the Fizeau drag after we take the nonzero fixed angular velocity $\Omega$ into account. The first few energy eigenstates are shown as Fig.\ref{fig.3}. The left (right) part shows the transitions when we drive the system from the left (right) side. It shows that when the light is resonantly coupled to the transition $|\psi_0\rangle \rightarrow |\psi
_1\rangle  $ by driving the device from its left side, the transition $|\psi_1\rangle \rightarrow |\psi _{2 \pm}\rangle$ are detuned by $\pm\sqrt{2} \hbar g$ so that it will be suppressed. However, there will be a two-photon resonance when we drive the cavity from the right side with the same laser frequency given the prerequisite that the angular velocity $\Omega$ is proper so that $\Delta _F=\frac{\sqrt{2}}{4}g$ ($\Delta _F=-\frac{\sqrt{2}}{4}g$) when we drive the cavity from the left (right) side. This means for the fundamental mode, the absorption of the first photon favors also that of the second photon, resulting in the transition $|1 \, 0 \rangle \rightarrow \frac{\sqrt{2}}{2}(|0 \, 1\rangle +|2\, 0\rangle )$, i.e., photon-induced tunneling (PIT).  

We assume to pump the fundamental cavity mode with the external classical light of frequency $\omega_L$, then by applying the operator $U=e^{i \omega _L t (a^ \dagger a +2 b^ \dagger b)}$, the effective Hamiltonian of the system can be written in a frame rotating with respect
to the laser frequency $\omega _L$ as
\begin{equation}
\begin{aligned}
    H_{eff}=&\hbar (\Delta  +\Delta _F ) a^\dagger a+ 2 \hbar (\Delta + \Delta _F) b^ \dagger b \\
    & +\hbar g (b {a^ \dagger} ^2+b^ \dagger a^2)+\hbar F (a+a^ \dagger)
    \label{3}
\end{aligned}
\end{equation}
where $\Delta =\omega _1-\omega _L$ is the detuning of the fundamental mode from the driving laser frequency and $F=\sqrt{\frac{2\kappa _1 P}{\hbar \omega_L}}$ denotes the driving strength  with cavity loss rate of the fundamental mode $\kappa_1$ and driving power $P$. Losses of the system can be described within a quantum master equation \cite{JOHANSSON20121760,JOHANSSON20131234}
\begin{equation}
    \frac{d \rho}{dt}=\frac{[H_{eff}, \rho]}{i\hbar}+\mathcal{L}[a](\rho)+\mathcal{L}[b](\rho)
    \label{4}
\end{equation}
where the Hamiltonian $H_{eff}$ is given by Eq.\eqref{3}, $\rho$ is the rotated density matrix and $\mathcal{L}[a](\rho)=\frac{\kappa}{2}(2a\rho a^ \dagger -a^ \dagger a \rho -\rho a^ \dagger a)$ and $\mathcal{L}[b](\rho)=\frac{\kappa}{2}(2b\rho b^ \dagger -b^ \dagger b \rho -\rho b^ \dagger b)$ are the Lindblad terms accounting for losses to the environment. The statistic properties of the photons for this nonlinear quantum system can be described by the second-order correlation function, defined as \cite{doi:10.1119/1.19344}
\begin{equation}
    g^{(2)}_{aa}(0)=\frac{<{a^\dagger}^2a^2>}{<a^\dagger a>^2}=\frac{Tr\{ {a^\dagger}^2a^2\rho_{ss}\}}{Tr^2\{a^\dagger a\rho_{ss}\}}
    \label{5}
\end{equation}
\begin{equation}
    g^{(2)}_{bb}(0)=\frac{<{b^\dagger}^2b^2>}{<b^\dagger b>^2}=\frac{Tr\{ {b^\dagger}^2b^2\rho_{ss}\}}{Tr^2\{b^\dagger b\rho_{ss}\}}
    \label{6}
\end{equation}
where $\rho_{ss}$ is the steady-state solution of Eq.\eqref{4} by setting the time derivative $\frac{d\rho}{dt}=0$. In the following, we numerically calculate the quantities of Eq.\eqref{5} and Eq.\eqref{6} by assuming realistic parameters for state-of-the-art nonlinear toroid cavity. 

\section{\label{sec:level3}Results and discussions}
For the study of PB, we assume that the driving light is extremely weak so we can restrict the photons within the low-excitation subspace. Then the second-order correlation functions can be calculated by solving the
master equation in Eq.\eqref{4} numerically.
\begin{figure}[! htpb]
    \centering\includegraphics[width=0.45\textwidth]{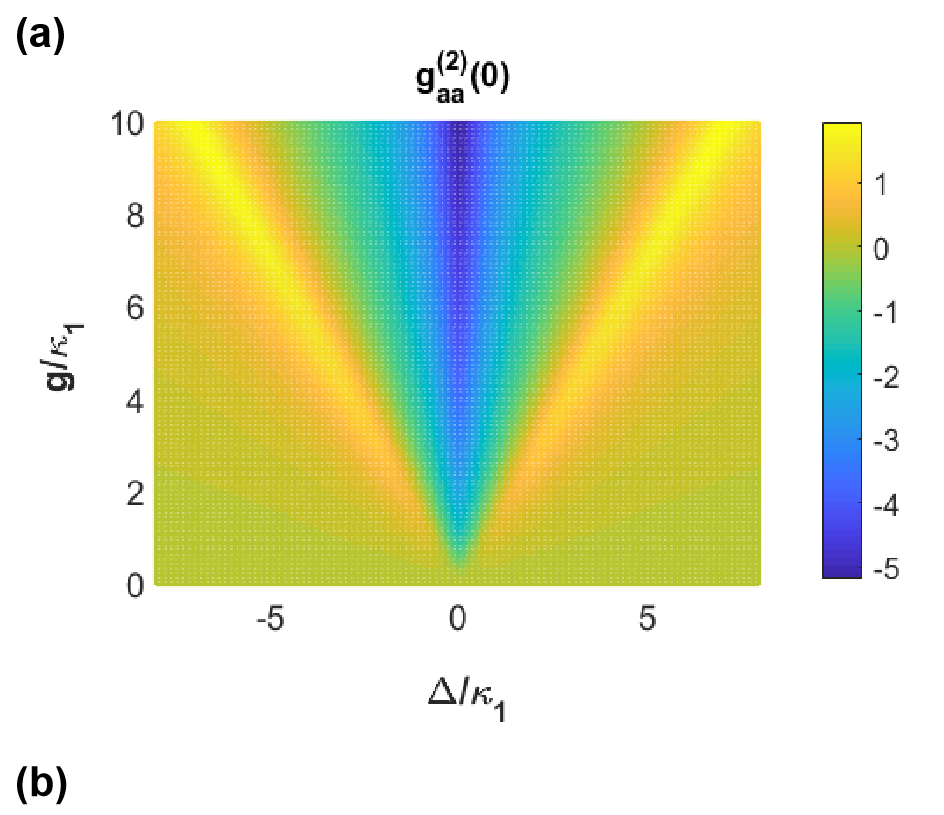}
    \centering\includegraphics[width=0.45\textwidth]{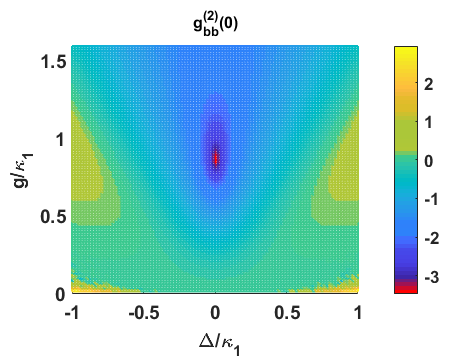}
    \caption{(a)The second-order correlation function at zero time delay for the fundamental mode $g^{(2)}_{aa}(0)$ as a function of the detuning of the fundamental mode from the driving laser frequency $\Delta$ and the hopping interaction $g$ between two modes caused by the second-order nonlinearity. (b)The second-order correlation function at zero time delay for the second-harmonic mode $g^{(2)}_{bb}(0)$ as a function of the detuning $\Delta$ and the hopping interaction $g$. Other parameters in both calculations are given by $\kappa_2=\kappa_2, F=0.05\kappa_1$ and the fixed angular velocity is 0.}
    \label{fig.4}
\end{figure}

\begin{figure}[! htpb]
    \centering\includegraphics[width=0.45\textwidth]{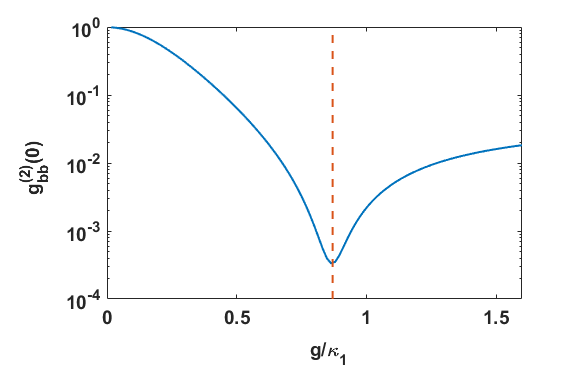}
     \caption{ The second-order correlation function at zero time delay for the second-harmonic mode $g^{(2)}_{bb}(0)$ versus the  hopping  interaction $g$ between two modes when the detuning $\Delta=0$. Other parameters are chosen as the same as we used before.}
    \label{fig.5}
\end{figure}

In Fig.\ref{fig.4}(a)(b), the second-order correlation function at zero time delay for the fundamental mode $g^{(2)}_{aa}(0)$ and the second-harmonic mode $g^{(2)}_{bb}(0)$ when the angular velocity $\Omega =0$ are shown on a logarithmic color scale plot as a function of the detuning of the fundamental mode from the driving laser frequency $\Delta$ and the hopping interaction between the two modes $g$, respectively.  In these calculations, we assumed similar cavity losses for the fundamental and second-harmonic modes, i.e., $\kappa_1=\kappa_2$. The experimentally accessible weak light strength is chosen as $F=0.05\kappa_1$. We can see in Fig.\ref{fig.4}(a) that the optimal antibunching occurs at $\Delta =0$, for the laser frequency exactly tuned with the fundamental cavity mode as the transition from $|\psi_0 \rangle$ to $|\psi_1 \rangle$ is enhanced with resonant photon absorption. This is perfectly matched with the schematic level diagram in Fig.\ref{fig.2}. Meanwhile, it is obvious that the PB gets better with the enhancement of the hopping interaction $g$ as the transition from $|\psi_1 \rangle$ to $|\psi_{2 \pm} \rangle$ is blocked for detuning $\sqrt{2}\hbar g$ which gets larger with increasing $g$. However, we can see in Fig.\ref{fig.4}(b), different from the monotone decreasing of $g^{(2)}_{aa}(0)$ according to the enhancement of $g$, there is an abnormal increase of the PB for the second-harmonic mode which appears when the hopping interaction $g$ is still small (red area). To show it more clearly, we set the detuning as $0$ and other parameters as the same then plot the second-order correlation function for the second-harmonic mode $g^{(2)}_{bb}(0)$ as a function of $g$ as show in Fig.\ref{fig.5}, from which shows a dip of the $g^{(2)}_{bb}(0)$ with a small value of $g \approx 0.867\kappa_1$. This abnormal phenomenon is the unconventional PB can be explained by the destructive interference between the two paths for two-photon excitation in the second-harmonic mode as we mentioned before. 

To examine the origin of this abnormal PB appearing with a weak second-order nonlinearity, we derive the optimal value of $g$ for the strongest PB under the resonant driving condition by following the method given in Ref.\cite{PhysRevA.83.021802} as follows. The average photon numbers in both modes remain small as the driving light is weak. Thus we can truncate the Hilbert space of the system by setting the cut-off occupations in the fundamental mode as four and that in the second-harmonic mode as two, i.e., $\langle a^\dagger a \rangle \leq 4 , \, \langle b^\dagger b \rangle \leq 2$, as the generation of a single photon with the second-harmonic frequency needs annihilating two photons with the fundamental frequency. Then we can expand the wave-function $|\psi (t) \rangle $ in the Fock-state basis as
\begin{equation}
    \begin{aligned}
        | \psi (t)  \rangle =& C_{00}(t)|0,0 \rangle + C_{10}(t)|1,0 \rangle + C_{20}(t)|2,0 \rangle \\ &+C_{30}(t)|3,0 \rangle  +C_{40}(t)|4,0 \rangle \\
        & +C_{01}(t)|0,1 \rangle +C_{11}(t)|1,1 \rangle \\
        & +C_{21}(t)|2,1 \rangle +C_{02}(t)|0,2 \rangle 
        \label{7}
    \end{aligned}
\end{equation}
with initial probability amplitudes satisfy $C_{00}(0)$ $\approx$ $1$ $ \gg$ $C_{10}(0)$ $\gg$ $C_{20}(0)$, $C_{01}(0)$ $\gg$ $C_{30}(0)$, $C_{11}(0)$ $\gg $ $C_{40}(0)$, $C_{21}(0)$, $C_{02}(0)$. The optical decay can be included in the Hamiltonian $H'= H_{eff}-i \hbar \frac{\kappa_1}{2}a^\dagger a-i \hbar \frac{\kappa_2}{2}b^\dagger b$ according to the quantum-trajectory method given in Ref.\cite{RevModPhys.70.101}. By substituting Eq.(\ref{7}) into $schr\ddot{o}dinger's$ equation $H' |\psi \rangle=i \hbar \dot{|\psi \rangle}$ with $\Delta=\Delta_F=0$, we have the following equations of motion
\begin{equation}
      i \hbar | \dot{C_{00}} (t)\rangle = C_{10}(t) \hbar F,
\label{8}
\end{equation}
\begin{equation}
    i \hbar | \dot{C_{10}} (t)\rangle =C_{00}(t) \hbar F +\underline{ \sqrt{2} C_{20}(t) \hbar F}-i C_{10}(t) \hbar \frac{\kappa_1}{2},
\end{equation}
\begin{equation}
    \begin{aligned}
    i \hbar | \dot{C_{20}} (t)\rangle &=\sqrt{2} C_{01}(t) \hbar g +\sqrt{2} C_{10}(t) \hbar F \\ &+\underline{ \sqrt{3} C_{30}(t) \hbar F} -i C_{20}(t)\hbar \kappa_1,
  \end{aligned}
\end{equation}
\begin{equation}
    \begin{aligned}   
    i \hbar | \dot{C_{30}} (t)\rangle &=\sqrt{6} C_{11}(t) \hbar g+\sqrt{3} C_{20}(t) \hbar F\\ &+\underline{2 C_{40}(t) \hbar F }-i C_{30}(t) \frac{3}{2}\hbar \kappa_1,
    \end{aligned}
\end{equation}
\begin{equation}
    \begin{aligned} 
    i \hbar | \dot{C_{40}} (t)\rangle =2\sqrt{3} C_{21}(t) \hbar g+2 C_{30}(t) \hbar F-iC_{40}(t) 2\hbar \kappa_1,
    \end{aligned}
\end{equation}
\begin{equation}
    \begin{aligned} 
    i \hbar | \dot{C_{01}}(t) \rangle =\sqrt{2} C_{20}(t) \hbar g+C_{11}(t) \hbar F-i C_{01}(t) \hbar \frac{\kappa_2}{2},
     \end{aligned}
\end{equation}
\begin{equation}
    \begin{aligned}
    i \hbar | \dot{C_{11}} (t)\rangle &=\sqrt{6} C_{30}(t) \hbar g+C_{01} (t)\hbar F+\underline{ \sqrt{2} C_{21}(t) \hbar F }\\ & -i C_{11}(t) \hbar \frac{\kappa_1}{2}-i C_{11}(t) \hbar \frac{\kappa_2}{2},
     \end{aligned}
\end{equation}
\begin{equation}
    \begin{aligned}
    i \hbar | \dot{C_{21}} (t)\rangle &=2C_{02}(t) \hbar g+2 \sqrt{3} C_{40} (t)\hbar g+\sqrt{2} C_{11}(t) \hbar F \\ & -i C_{21}(t) \hbar \kappa_1-i C_{21}(t) \hbar \frac{\kappa_2}{2},
     \end{aligned}
\end{equation}
\begin{equation}
   i \hbar | \dot{C_{02}} (t)\rangle =2 C_{21}(t) \hbar g-i C_{02}(t) \hbar \kappa_2.
   \label{16}
\end{equation} 
in which the underlined terms have the subleading-order coefficients of $F$ and thus can be neglected. We can derive the steady-state solution $C_{02}( \infty)$ from Eqs.(\ref{8})-(\ref{16}), then by setting $C_{02}( \infty)=0$, the optimal condition for the strongest PB of the second-harmonic mode is obtained as
\begin{equation}
    g=\frac{\sqrt{4F^2+(2\kappa_1+\kappa_2)(\kappa_1+\kappa_2)}}{2\sqrt{2}}
    \label{17}
\end{equation}
That is $g=0.867\kappa_1$ when taking $\kappa_2=\kappa_1$ and $F=0.05\kappa_1$ into the calculation, which is consistent with the numerical result shown in Fig.\ref{fig.5}. 

\begin{figure}[! htpb]
      \centering\includegraphics[width=0.45\textwidth]{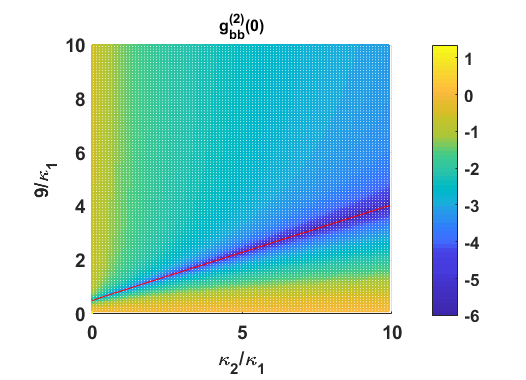}
     \caption{ The second-order correlation function at zero time delay for the second-harmonic mode $g^{(2)}_{bb}(0)$ is drawn as a function of the cavity loss rate of the second-harmonic mode $\kappa _2$ and the hopping interaction $g$ between the two modes. The detuning $\Delta$ and the fixed angular velocity have been set as 0. The driving strength is chosen as $0.05\kappa_1$. The red curve represents the optimal condition for the PB according to the analytical calculation.}
    \label{fig.6}
\end{figure}
By setting $\Delta=\Delta_F=0$ and $F=0.05\kappa_1$, the dependence of the second-order correlation function for the second-harmonic mode is checked against the cavity loss rate of the second-harmonic mode $\kappa_2$ and the second nonlinearity $g$ in Fig.\ref{fig.6}. The red curve shows the optimal values of $g$ versus $\kappa_2$ for the strongest PB according to Eq.(\ref{17}) with $F$ chosen as $0.05\kappa_1$ as well. We can see the red curve is perfectly agrees with the numerical results. This suggests again that the strong PB appearing with a lower $g$ arising from the destructive interference between the two paths for two-photon excitation in the second-harmonic mode.

\begin{figure}[! htpb]
    \centering
    \includegraphics[width=0.45\textwidth,height=2.9in]{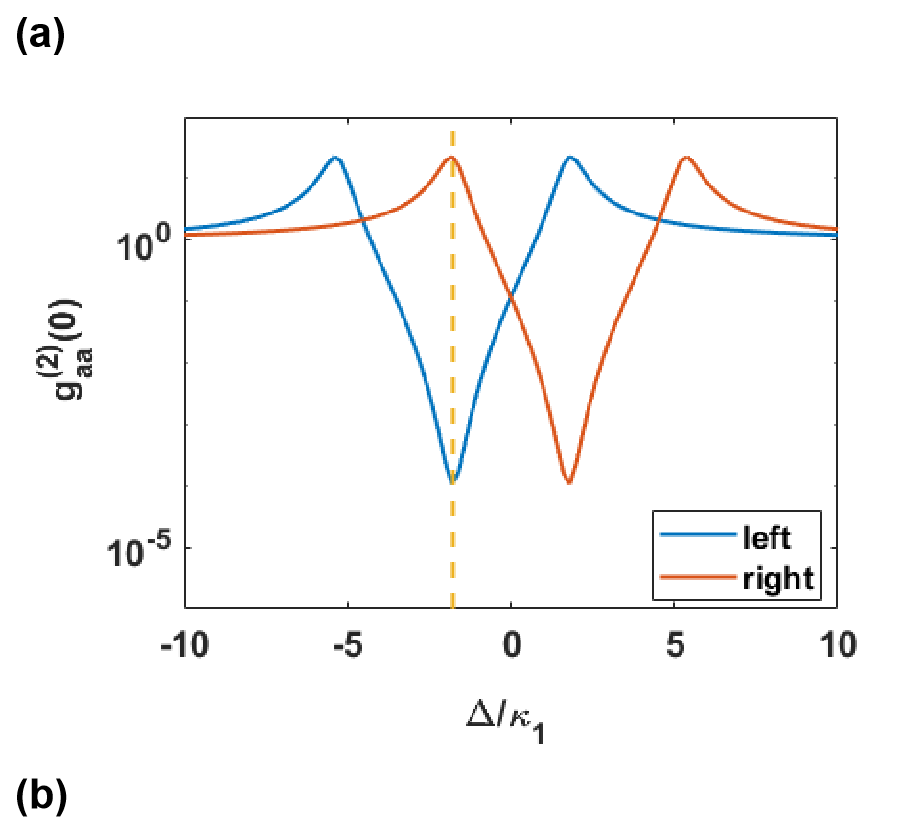}
    \centering
    \includegraphics[width=0.45\textwidth,height=2.5in]{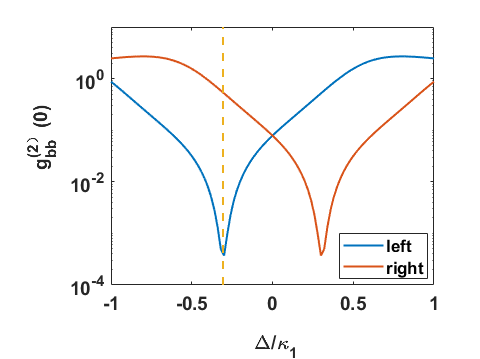}
    \caption{The second-order correlation function at zero time delay for the fundamental mode $g^{(2)}_{aa}(0)$ (a) and that for the second-harmonic mode $g^{(2)}_{bb}(0)$ (b) are drawn as a function of the detuning $\Delta$ when the fixed angular velocity meets $\Delta_{F}=\pm \frac{nr \Omega \omega _1}{c}(1-\frac{1}{n^2}-\frac{\lambda}{n}\frac{dn}{d \lambda})=\pm \frac{\sqrt{2}}{4}g$. The blue (red) curve represents we driving the system from the left (right) side, and the orange dotted line refer to $\Delta=-\Delta_F$ when we drive the system from the left side. Other parameters are given by $\kappa_2=\kappa_1$, $F=0.05\kappa_1$, and $g=5\kappa_1$ in (a), $g=\frac{\sqrt{4F^2+(2\kappa_1+\kappa_2)(\kappa_1+\kappa_2)}}{2\sqrt{2}}=0.867\kappa_1$ in (b).}
    \label{fig.7}
\end{figure}
\begin{figure}[! htpb]
    \centering
    \includegraphics[width=0.45\textwidth]{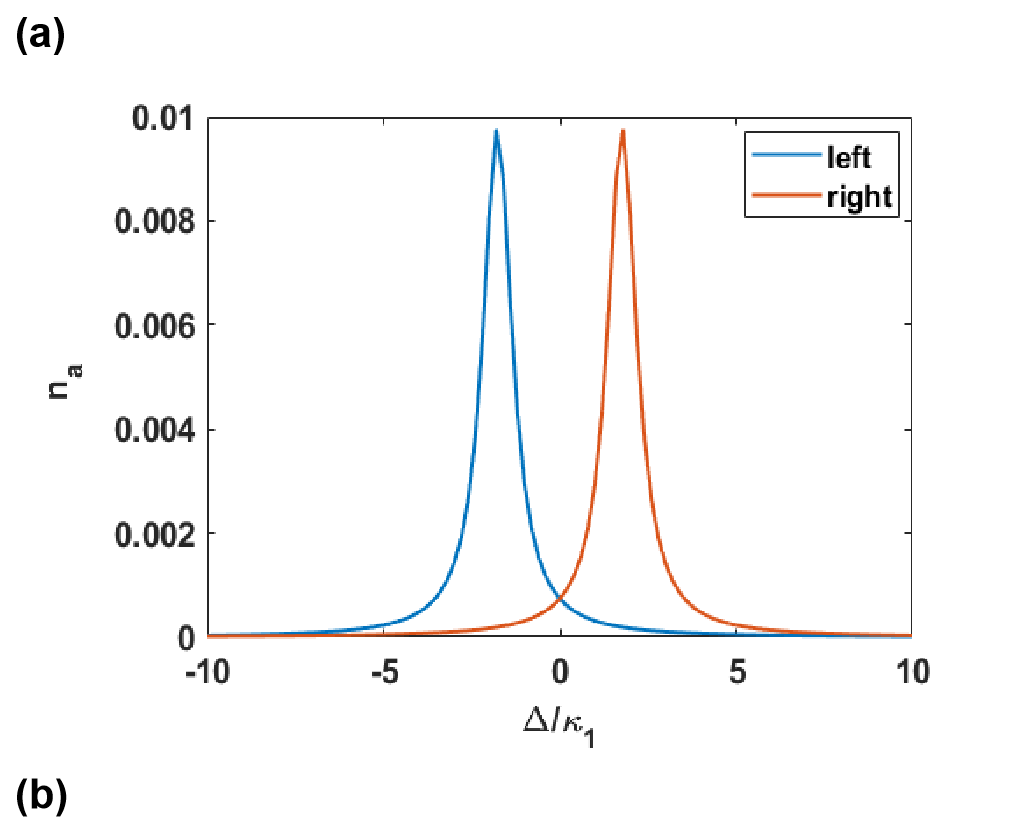}
    \centering
    \includegraphics[width=0.45\textwidth]{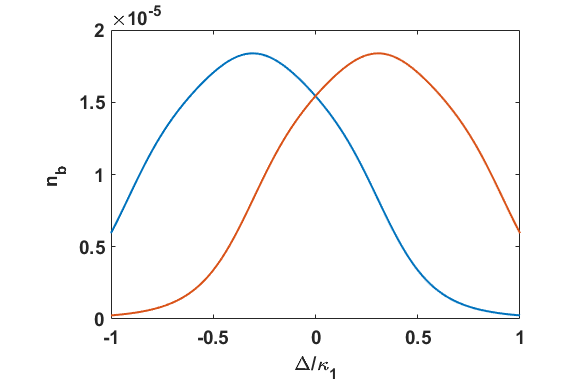}
    \caption{The photon number with the fundamental frequency $n_a$ (a) and that with the second-harmonic frequency $n_b$ (b) are drawn as a function of the detuning $\Delta$ when the fixed angular velocity meets $\Delta_{F}=\pm \frac{nr \Omega \omega _1}{c}(1-\frac{1}{n^2}-\frac{\lambda}{n}\frac{dn}{d \lambda})=\pm \frac{\sqrt{2}}{4}g$. The blue (red) curve represents we driving the system from the left (right) side. Other parameters are given by $\kappa_2=\kappa_1$, $F=0.05\kappa_1$ and $g=5\kappa_1$ in (a), $g=\frac{\sqrt{4F^2+(2\kappa_1+\kappa_2)(\kappa_1+\kappa_2)}}{2\sqrt{2}}=0.867\kappa_1$ in (b).}
    \label{fig.8}
\end{figure}
Finally, we check the second-order correlation function while there are energy-level transitions caused by the Fizeau drag. We can get $\Delta_F=\pm \frac{\sqrt{2}}{4}g$ by driving the system from different sides given condition that the fixed angular velocity is a proper nonzero value $\Omega$ satisfies $\frac{nr \Omega \omega _1}{c}(1-\frac{1}{n^2}-\frac{\lambda}{n}\frac{dn}{d \lambda})= \frac{\sqrt{2}}{4}g$. The other parameters are chosen as $\kappa_2=\kappa_1$ and $F=0.05\kappa_1$, and for producing a strong antibunching, we set $g=5\kappa_1$ in (a), $g=\frac{\sqrt{4F^2+(2\kappa_1+\kappa_2)(\kappa_1+\kappa_2)}}{2\sqrt{2}}=0.867\kappa_1$ in (b). The second-order correlation function of the fundamental and second-harmonic modes versus the detuning $\Delta$ are shown in Fig.\ref{fig.7}(a)(b). Both the blue (red) curves represent that we drive the system from the left (right) side, and the yellow dotted lines refer to $\Delta=-|\Delta_F|$. It is apparent that the second-order correlation functions of both modes reach their dips at $\Delta=-\Delta_F$ when we drive the system from the left side. Moreover, in Fig.\ref{fig.7}(a), the red curve reaches its peak at the exact point where the blue curve reaches its dip, and the value of $g^{(2)}_{aa}(0)$ at this point is lager than 1, which means for the fundamental mode, when we drive the system from the right side, there happens the photon-induced tunneling with the optimal detuning for the PB when we drive the system from the left side. This phenomenon is perfectly matched with the schematic level diagram in \ref{fig.3} and is exactly what we expected for the nonreciprocal feature of the PB. However, in Fig.\ref{fig.7}(b), there is no peak for $g^{(2)}_{bb}(0)$ at $\Delta=-|\Delta_F|$ when we drive the system from the right side. This is because the PB in the second-harmonic mode arising from the destructive interference which is not affected by the energy-level shifts instead of the anharmonic energy-level structure.

Mean photon numbers in the fundamental mode $n_a$ and the second-harmonic mode $n_b$ are shown in Fig.\ref{fig.8}(a)(b) as a function of the detuning $\Delta$. Other parameters are the same as we used in Fig\ref{fig.7}. In Fig.\ref{fig.8}(a), we can see the nonreciprocity clearly at the point $\Delta=-|\Delta_F|$ from the different efficiencies for single-photon generation in two modes. $n_a$ (a) is, on average, much larger than $n_b$ (b), which is because the generation of a single photon in the second-harmonic mode needs annihilating two photons in the fundamental mode just like we have discussed before.

\section{\label{sec:level4}Conclusion}
In conclusion, we have proposed a physical model of rotating nanostructured toroid cavity made of second nonlinear materials for the application of achieving nonreciprocal PB at the output of the system under continuous weak driving condition. We show that the PB phenomenon in the fundamental mode happens based on the anharmonicity in energy eigenvalues of the system and gets stronger with a higher second-order nonlinearity. However, the PB phenomenon in the second-harmonic mode happens based on the destructive interference between different pathways for the two-photon excitation. The strongest PB in this mode happens with an optimal condition between the hopping interaction and the cavity loss rate of the second-harmonic mode. There is nonreciprocity phenomenon generated from the Fizeau shift but only happens in the fundamental mode not the second-harmonic mode because the energy-level shifts will not affect the destructive interference between two pathways for the two-photon
excitation.

\section*{Acknowledgement}
This work was supported by
the National Natural Science Foundation of China (Nos. 11574092, 61775062,
61378012, 91121023); the National Basic Research Program of China (No.
2013CB921804).


\begin{thebibliography}{99}

\bibitem{PhysRevA.61.011801} 
M. J. Werner and A. Imamo\ifmmode \bar{g}\else \={g}\fi{}lu, Phys. Rev. A 
\textbf{61}, 011801
(1999).

\bibitem{PhysRevLett.109.013603}K. Stannigel, P. Komar, S. J. M. Habraken, S. D. Bennett, M. D. Lukin, P. Zoller, and P. Rabl, Phys. Rev.
Lett. \textbf{109}, 013603 (2012).

\bibitem{10.1038/35005001}C. H. Bennett and D. P. DiVincenzo, Nature \textbf{404},  247
(2000).

\bibitem{Buluta_2011}I. Buluta, S. Ashhab, and F. Nori, Reports on Progress
in Physics \textbf{74},  104401
(2011).

\bibitem{RevModPhys.81.1301} V. Scarani, H. Bechmann-Pasquinucci, N. J. Cerf,
M. Du\ifmmode \check{s}\else \v{s}\fi{}ek, N. L\"utkenhaus, and M. Peev, Rev. Mod.
Phys. \textbf{81}, 1301 (2009).

\bibitem{10.1038/nphoton.2009.229}  J. L. O'Brien, A. Furusawa, and J. Vukovi, Nature Photonics \textbf{3}, 687 (2009)

\bibitem{10.1038/nature07127}H. J. Kimble, Nature \textbf{453}, 1023 (2008)

\bibitem{PhysRevLett.106.243601} C. Lang, D. Bozyigit, C. Eichler, L. Steffen, J. M. Fink,
A. A. Abdumalikov, M. Baur, S. Filipp, M. P. da Silva,
A. Blais, and A. Wallraff, Phys. Rev. Lett. \textbf{106}, 243601 (2011).

\bibitem{PhysRevLett.107.053602}  A. J. Hoffman, S. J. Srinivasan, S. Schmidt, L. Spietz,
J. Aumentado, H. E. T\"ureci, and A. A. Houck, Phys.
Rev. Lett. \textbf{107}, 053602 (2011).


\bibitem{PhysRevA.89.043818} Y.-x. Liu, X.-W. Xu, A. Miranowicz, and F. Nori, Phys.
Rev. A \textbf{89}, 043818 (2014).

\bibitem{PhysRevLett.94.053604} M. Hennrich, A. Kuhn, and G. Rempe, Phys. Rev. Lett.
\textbf{94}, 053604 (2005).

\bibitem{PhysRevLett.96.093603} W. Choi, J.-H. Lee, K. An, C. Fang-Yen, R. R. Dasari,
and M. S. Feld, Phys. Rev. Lett. \textbf{96}, 093603 (2006).

\bibitem{PhysRevLett.107.063601}  P. Rabl, Phys. Rev. Lett. \textbf{107}, 063601 (2011).

\bibitem{PhysRevA.92.033806}  H. Wang, X. Gu, Y.-x. Liu, A. Miranowicz, and F. Nori,
Phys. Rev. A \textbf{92}, 033806 (2015).

\bibitem{PhysRevA.76.031805}  D. G. Angelakis, M. F. Santos, and S. Bose, Phys. Rev.
A \textbf{76}, 031805 (2007).

\bibitem{PhysRevA.94.013815} H. Flayac and V. Savona, Phys. Rev. A \textbf{94}, 013815
(2016).

\bibitem{PhysRevA.91.063808} H. Z. Shen, Y. H. Zhou, and X. X. Yi, Phys. Rev. A \textbf{91},
063808 (2015).

\bibitem{PhysRevLett.107.223601} H. Zheng, D. J. Gauthier, and H. U. Baranger, Phys.
Rev. Lett. \textbf{107}, 223601 (2011).

\bibitem{10.1038/nature11361}  T. Peyronel, O. Firstenberg, Q.-Y. Liang, S. Hofferberth,
A. V. Gorshkov, T. Pohl, M. D. Lukin, and V. Vuleti,
Nature \textbf{488}, 57 (2012).

\bibitem{PhysRevA.96.053810} H. Flayac and V. Savona, Phys. Rev. A \textbf{96}, 053810
(2017).

\bibitem{PhysRevLett.121.153601}R. Huang, A. Miranowicz, J.-Q. Liao, F. Nori, and
H. Jing, Phys. Rev. Lett. \textbf{121}, 153601 (2018).

\bibitem{Sounas2017}D. L. Sounas and A. Al, Nature Photonics \textbf{11}, 774 (2017).

\bibitem{PhysRevLett.120.043901}H. Ramezani, P. K. Jha, Y. Wang, and X. Zhang, Phys.
Rev. Lett. \textbf{120}, 043901 (2018).

\bibitem{PhysRevLett.110.093901}D.-W. Wang, H.-T. Zhou, M.-J. Guo, J.-X. Zhang, J. Evers, and S.-Y. Zhu, Phys. Rev. Lett. \textbf{110}, 093901 (2013).

\bibitem{Fan447} L. Fan, J. Wang, L. T. Varghese,
H. Shen, B. Niu, Y. Xuan, A. M. Weiner, and M. Qi, Science \textbf{335}, 447 (2012).

\bibitem{PhysRevLett.118.033901}Q.-T. Cao, H. Wang, C.-H. Dong, H. Jing, R.-S. Liu,
X. Chen, L. Ge, Q. Gong, and Y.-F. Xiao, Phys. Rev.
Lett. \textbf{118}, 033901 (2017).

\bibitem{PhysRevLett.102.213903}
S. Manipatruni, J. T. Robinson, and M. Lipson, Phys.
Rev. Lett. \textbf{102}, 213903 (2009)

\bibitem{10.1038/nphoton.2016.161}Z. Shen, Y.-L. Zhang, Y. Chen, C.-L. Zou, Y.-F. Xiao, X.-
B. Zou, F.-W. Sun, G.-C. Guo, and C.-H. Dong, Nature
Photonics \textbf{10}, 657 (2016).

\bibitem{PhysRevLett.110.234101}N. Bender, S. Factor, J. D. Bodyfelt, H. Ramezani, D. N.
Christodoulides, F. M. Ellis, and T. Kottos, Phys. Rev.
Lett. \textbf{110}, 234101 (2013)

\bibitem{nphoton.2014.133} L. Chang, X. Jiang, S. Hua, C. Yang, J. Wen, L. Jiang,
G. Li, G. Wang,  and M. Xiao, Nature Photonics
\textbf{8}
, 524 (2014).

\bibitem{PhysRevA.89.031803} D. Gerace and V. Savona, Phys. Rev. A \textbf{89}, 031803
(2014).

\bibitem{PhysRevLett.96.057405} W. T. M. Irvine, K. Hennessy, and D. Bouwmeester,
Phys. Rev. Lett. \textbf{96}, 057405 (2006).

\bibitem{Malykin_2000} G. B. Malykin, Physics-Uspekhi \textbf{43}, 1229 (2000).

\bibitem{PhysRevB.87.235319} A. Majumdar and D. Gerace, Phys. Rev. B \textbf{87}, 235319
(2013).

\bibitem{JOHANSSON20121760} J. Johansson, P. Nation, and F. Nori, Computer Physics
Communications \textbf{183}, 1760 (2012).

\bibitem{JOHANSSON20131234} J. Johansson, P. Nation, and F. Nori, Computer Physics
Communications \textbf{184}, 1234 (2013).

\bibitem{doi:10.1119/1.19344} M. O. Scully and M. S. Zubairy, American Journal of
Physics \textbf{67}, 648 (1999).

\bibitem{PhysRevA.83.021802} M. Bamba, A. Imamo\ifmmode \breve{g}\else \u{g}\fi{}lu, I. Carusotto, and C. Ciuti,
Phys. Rev. A \textbf{83}, 021802 (2011).

\bibitem{RevModPhys.70.101} M. B. Plenio and P. L. Knight, Rev. Mod. Phys. \textbf{70}, 101
(1998).


\end{thebibliography}
\end{document}